*Research article*

# Investigating cerebral anomalies in preterm infants and associated risk factors with MRI at term-equivalent age

**Short title:** Cerebral anomalies of prematurity and associated risk factors

**Authors:** Nicolas Elbaz[1], Valérie Biran[2,3], Chloé Ghozland[2], Laurie Devisscher[3,4], Aline Gonzales-Carpinteiro[3,4], Aurélie Bourmaud[5], Monique Elmaleh-Bergès[1,3], Lucie Hertz-Pannier[3,4], Yann Leprince[4], Alice Frérot[2], Alice Héneau[2], Jessica Dubois[3,4]*, Marianne Alison[1,3]*

**Affiliations:**

[1] Service d'Imagerie Pédiatrique, Hôpital Robert-Debré, AP-HP Paris, Université Paris-Cité

[2] Service de réanimation et pédiatrie néonatales, Hôpital Robert-Debré, AP-HP Paris, Université Paris-Cité

[3] Université Paris-Cité, Inserm U1141, NeuroDiderot, F-75019 Paris

[4] Université Paris-Saclay, CEA, NeuroSpin, UNIACT, U1141, F-91191 Gif-sur-Yvette

[5] Unité d'épidémiologie clinique, CHU Robert-Debré, AP-HP Paris, Inserm CIC 1426 and Université Paris-Cité

* co-last authors

**Corresponding Author:** Jessica Dubois

**Category of study:** Clinical study

**Data availability:** Derived data generated will be shared on reasonable request to the corresponding author.

**Funding/Support:** This work was supported by the French National Agency for Research (ANR-20-CE17-0014, PremaLocom project), the Fondation Médisite (FdF-18-00092867), the IdEx Université de Paris (ANR-18-IDEX-0001) and the French government as part of the France 2030 programme (ANR-23-IAHU-0010, IHU Robert-Debré du Cerveau de l'Enfant).

**Abbreviations**

ADC – apparent diffusion coefficient

BPD – broncho-pulmonary dysplasia

CI – Confidence interval

DEHSI – diffuse excessive high signal intensity

FOV – field of view

GA – gestational age

GM – grey matter

MRI – magnetic resonance imaging

NDD – neurodevelopmental disorders

NEC – necrotizing enterocolitis

NICU – neonatal intensive care unit

OR – Odds-ratio

PMA – post-menstrual age

SGA – small for gestational age

SWI – susceptibility-weighted imaging

WM – white matter

# Abstract

**Background:** Being born very or extreme preterm is a major source of cerebral anomalies and neurodevelopmental disorders, whose occurrence depends on many perinatal factors. A better understanding of these factors could be provided by cerebral Magnetic Resonance Imaging (MRI) at term-equivalent age (TEA).

**Objective:** To investigate, through cerebral TEA-MRIs, the relationship between the main perinatal factors, the occurrence of cerebral anomalies, and cerebral volumetry.

**Methods:** We assembled a cohort of preterm babies born before 32 weeks of gestation who underwent a cerebral TEA-MRI. We assessed cerebral anomalies using a radiological scoring system – the Kidokoro scoring – and performed cerebral volumetry. We investigated the relationships between 9 clinical factors (birth characteristics, resuscitation treatments…), Kidokoro scores, and brain volumes.

**Results:** Among 110 preterms who underwent a cerebral MRI at TEA, only 6% suffered moderate-to-severe brain anomalies. We identified mechanical ventilation as a risk factor for cerebral anomalies (adjusted Odds-Ratio aOR = 4.6, 95% Confidence Interval CI [1.7-12.8]) and prolonged parenteral nutrition as a protective factor for white matter anomalies (aOR = 0.2, 95%CI [0.1-0.8]). Mechanical ventilation ($p = 0.01$) and being born small for gestational age ($p < 0.001$) were risk factors for the reduction of cerebral volumes. An increase in brain lesion severity was associated with decreased cerebral volumes ($p = 0.016$).

**Conclusion:** Our study highlights the importance of treatment-related perinatal factors on the occurrence of cerebral anomalies in very and extreme preterms, and the interest in using both qualitative (Kidokoro scoring) and quantitative (volumetry) MRI-tools.

# Introduction

Prematurity is a frequent phenomenon, occurring in approximately 10% of all births worldwide [1]. It is one of the main causes for neurodevelopmental disorders (NDD), especially in very and extreme preterm births (occurring before 32 weeks of gestation) (2% of all births) [2] [3]. Those disorders imply different kinds of limitations [4], among which cerebral palsy [5] or autistic spectrum disorders [6], and therefore lead to different degrees of disability. Since NDDs often arise from prematurity-associated brain anomalies [7], it was stated that neuroimaging through Magnetic Resonance Imaging (MRI) at term-equivalent-age (TEA) may help predict the neurodevelopmental prognosis of the infant born preterm [8]. Indeed, brain MRI provides a large number of analytic tools, either qualitative or quantitative, based on both routine and advanced imaging techniques [9]. Such insights on the neurodevelopmental prognosis of preterms could allow for better risk stratification and optimized care strategies.

Additionally, the occurrence of brain anomalies in preterms and the ensuing disorders are themselves influenced by many perinatal clinical factors, that mainly represent risk factors [10]. Among them, pro-inflammatory factors play a major role. For example, prenatal chorioamnionitis can lead to reduced cerebral volumes [11] and later cognitive performances in children [12]. Postnatal necrotizing enterocolitis (NEC) has been associated with white matter lesions [13] and cerebral palsy [14]. A thorough knowledge of those clinical factors is thus essential to evaluate the prognosis of preterm neonates. Yet, the role of those factors in brain resilience or vulnerability is not fully understood.

To better understand the clinical factors affecting brain development in preterms during the perinatal period, we aimed at investigating the relationship between the main early clinical factors and the occurrence of cerebral anomalies in the very and extremely preterm infants: using MRI-scans at TEA, we searched first for an association between clinical factors and a global brain abnormality score developed for preterms by Kidokoro et al. (or "Kidokoro scoring") [15] and widely used by clinicians, and secondarily for an association between the same factors and measures of brain volumetry.

## Methods

### Population and study design

This is a retrospective cohort study of neonates admitted between April 2021 and July 2024 to tertiary neonatal intensive care units (NICU) right after birth in Paris area. Although preterm neonates were born in different maternities, TEA-MRI was performed on the same 3T MRI scanner at Robert-Debré Hospital, with a standardized protocol. We included all preterm neonates born at a gestational age below 32 weeks, that had undergone a TEA-MRI between 38 to 43 weeks of post-menstrual age (PMA). The requirements to undergo a TEA-MRI were either to be born extremely preterm (before 28 weeks), or to be born very preterm (between 28 to 32 weeks) with cerebral anomalies on cranial ultrasonography. We excluded the neonates with significant malformations or genetic/syndromic disorders.

This study was approved by the Ethic Committee of research in medical imaging, of the CERF (Collège des Enseignants en Radiologie de France), and the CPP (Comité de Protection des Personnes) Ile de France 3 (protocol DEVine, CEA 100 054). All procedures were carried out in accordance with the ethical principles established in the Declaration of Helsinki and local regulations regarding data protection.

### TEA-MRI

TEA-MRI was conducted on an Ingenia Philips Healthcare 3.0 Tesla MR system, using a multi-channel head coil, without sedation (spontaneous sleep was obtained after feeding).

MRI protocol consisted of a systematic multimodal evaluation, including:

- 2D T2-weighted imaging in coronal, axial and sagittal planes (echo time [TE] = 150 ms, repetition time [TR] = 5500 ms, flip angle = 90° ; in-plane FOV = 153 x 153 mm, reconstructed matrix = 192 x 192 and resolution = 0.8 x 0.8 mm²; slice thickness = 2.0 mm, 70, 60 and 60 slices respectively), duration 3'51'', 3'18'' and 3'18'' respectively
- A conventional sagittal 3D T1-weighted imaging (TE = 33 ms, TR = 500 ms, flip angle = 90°; FOV = 170 x 176 x 140 mm; reconstructed matrix 224 x 224 x 175 and resolution = 0.76 x 0.79 x 0.8 mm), duration 3'54''

- An axial 3D Susceptibility Weighted Imaging SWI (TE = 6 ms, TR = 40 ms, 6 echoes separated by 6ms, flip angle = 17°; FOV = 157 x 179 x 140 mm; reconstructed matrix = 196 x 224 x 93 and resolution 0.8 x 0.8 x 1.5 mm), 4'13''
- An axial diffusion-weighted imaging (with a b value of 0 and 1000 s/mm² for six directions of diffusion gradients) for computation of Apparent Diffusion Coefficient (ADC) (TE = 78 ms, TR shortest = 3330-3350 ms; FOV = 180 x 180 x 110 mm, reconstructed matrix = 288 x 288 x 54 and resolution = 0.625 x 0.625 x 2mm), duration 2'20''.

**Qualitative analysis of MRI: Kidokoro scoring**

Analysis of MRI was performed on a workstation (Carestream Vue PACS software) by two operators: a senior and a junior radiologist (MA and NE). In case of disagreement, a consensus was obtained.

Qualitative analysis was performed by the same operator (junior radiologist NE) using standardized radiological Kidokoro scoring system [15]. An abnormality score and degree of severity were thus obtained for each cerebral compartment: grey matter GM (cortex and central grey nuclei), white matter WM, cerebellum, and for the whole brain. Cortical abnormality was graded based on 3 variables: signal abnormality, delayed gyration and increased extracerebral space; while central grey nuclei abnormality was graded based on signal abnormality and gross volume reduction (evaluated by the deep GM area). GM score was classified as normal (0), mild (1), moderate (2), or severe (≥ 3).

The occurrence of cerebral WM injury was graded based on 6 variables: cystic lesion, focal signal abnormalities, delayed myelination, thinning of the corpus callosum, dilated lateral ventricles (we considered a ventricular dilatation when lateral ventricular diameter was above 10 mm on a coronal slice at the level of the atria) and gross reduction of WM volume (evaluated by the biparietal diameter). WM score was classified as normal (0-2), mild (3-4), moderate (5-6), or severe (≥ 7).

Cerebellum was also graded based on signal abnormality and gross volume reduction (evaluated by transverse cerebellar diameter). Cerebellum score was classified as normal (0), mild (1), moderate (2), and severe (≥ 3).

Finaly, a whole brain score was calculated as a sum of the three regional subscores and classified as normal (0-3), mild (4-7), moderate (8-11), and severe (≥ 12).

Presence of hemorrhagic lesions was analyzed according to Papile classification, adapted to MRI [16] [17]. Previous intraventricular hemorrhage was diagnosed when hemosiderin deposits were detected within the ventricles with SWI. Ventricular dilatation considered when lateral ventricular diameter measured at the level of the atrium on a T2-weighted coronal slice was above 10 mm. In the absence of ventricular dilatation, which could correspond to resolving post-hemorrhagic ventricular dilatation, lesions were classified as grade II. In case of persistent dilation, they were classified as grade III. When focal hemosiderin deposits were detected in the periventricular white matter with porencephalic cavity or focal white matter atrophy, which could correspond to venous infarction, lesions were classified as grade IV.

Diffusion-weighted imaging was further used to perform quantitative ADC measurement in right and left frontal and parietal WM. An increased ADC value according to normative values (above 1.7 mm$^2$/s) [18] was classified as "Diffuse excessive high signal intensity (DEHSI)".

For the Kidokoro scoring, a reproducibility study was performed on twenty subjects. These subjects were selected according to their degree of severity, in order to obtain a fair distribution of each category. A second evaluation was performed by the same operator at a distance from the first assessment, to estimate an intra-observer concordance coefficient.

**Quantitative analysis: cerebral volumetry**

Besides, cerebral volumetry was performed using state-of-the-art tools. We first reconstructed a T2-weighted volume with a 0.8 mm isotropic resolution, using the super-resolution tool NiftyMIC [19] based on at least two different orientations (coronal, axial and/or sagittal) without movement artefacts. The resulting images were used to segment the different brain tissues and compartments by coupling two automatic tools dedicated to the neonatal brain: DrawEM (Developing brain Region Annotation With Expectation-Maximization) [20] and iBEAT (infant Brain Extraction and Analysis Toolbox) [21]. As each segmentation showed its own qualities and shortcomings, they were combined (by masking and applying tools of mathematical morphology) to obtain the following compartments in the most reliable and reproducible way possible: cortical and central grey matter (combined in the following), sub-cortical and central white matter (combined in the following), cerebellum, brainstem, hippocampus and cerebro-spinal fluid (**Figure 1a**). Visual inspections by the same operator (NE) highlighted some significant errors and inaccuracies in about 25% of segmentations: each segmentation was assessed and received a grading ("perfect", "good", "acceptable", "bad" and "unusable"). The "bad" and "unusable" segmentations had to undergo manual corrections (by AGC and JD) and further inspection (by NE). The compartment volumes of grey matter, white matter and cerebellum were then quantified.

**Clinical characteristics**

For each patient, clinical data referring to birth (gestational age at birth, birth weight, sex) and the presence of severe neonatal morbidities (such as and bronchopulmonary dysplasia and necrotizing enterocolitis) were prospectively recorded into the medical chart.

Clinical data of interest were then collected retrospectively using the hospital electronic health record system. It should be noted that, in this retrospective study, the ethical authorization to collect and report clinical data only involved the infants with an MRI exam, but not those who died before term-equivalent age.

We inquired for 9 clinical factors, considered to be among the main neonatal morbidity risk factors: sex, monochorionic twin pregnancy, category of gestational age at birth, being small for gestational age at birth (SGA), bronchopulmonary dysplasia (BPD), use of mechanical invasive ventilation, necrotizing enterocolitis (NEC), use of prolonged parenteral nutrition, and perinatal sepsis.

The sex of the preterm was considered because moderate-to-severe NDDs are more frequent in preterm boys, suggestive of a "protective role" of female sex [22]. We considered specifically monochorionic twin pregnancies as the occurrence of brain lesions is higher compared with dichorionic ones, due to vascular complications [23]. Gestational age was estimated based on first trimester ultrasound, and three groups were considered: born below 26 weeks PMA (G1), between 26 to 28 weeks (G2), and between 28 to 32 weeks (G3). Infants were classified as small for gestational age at birth if they were born with a birth weight below the $10^{th}$ percentile on customized

AUDIPOG curves for male and female neonates [24]. BPD was defined as a requirement for oxygen-therapy beyond 36 weeks of PMA or for more than 28 days after birth. Mechanical invasive ventilation was only considered if it lasted strictly more than one day, to exclude infants who received it for short surgical procedures but to avoid an arbitrary threshold given the negative association observed between any additional day of ventilation and neurodevelopment [25]. NEC diagnosis was based on both clinical data and imaging according to Bell's classification [26], only stage ≥ II NEC were considered, whether perforated or non-perforated. Prolonged parenteral nutrition was defined as a continued parenteral nutrition for more than 21 days. Sepsis was defined as any infection requiring antibiotic treatment for longer than five days, whether a bacterial pathogen had been detected or not.

**Statistical analysis**

Data were analysed using the R 4.3.3 programming software (R Core Team 2024). Discrete and categorical variables (the nine clinical factors, Kidokoro scores and injury severity according to Kidokoro scoring, hemorrhagic lesions, DEHSI) were expressed as frequency and percentage. Continuous variables (PMA at TEA-MRI, cerebral volumes) were expressed as mean and standard deviation as appropriate. Hemorrhagic lesions and DEHSI were reported here for descriptive purposes only, and were not considered for regression analysis.

The relationship between Kidokoro injury severity and the nine clinical factors was explored for the whole brain and for each cerebral compartment (grey matter, white matter, and cerebellum) using a logistic regression analysis. A p-value of < 0.05 was taken as statistically significant.

The relationship between volumes, corresponding Kidokoro injury severity, and clinical factors was explored for the whole brain and for each compartment using a multivariate linear regression analysis, considering the post-menstrual age at TEA-MRI as a covariate. A p-value of < 0.05 was taken as statistically significant. Some trends for p < 0.1 are also reported.

# Results

## Descriptive results

A total of 110 very and extreme preterm infants underwent a TEA-MRI between 38 to 42 weeks+ 6 days of PMA (Mean (SD) PMA in weeks: 41.1 (0.73)). Among them, 54% (59/110) were males. The mean (SD) GA at birth was 27.4 (2) weeks of gestation, with 31% (34/110) born before 26 weeks, 35 % (39/110) between 26 and 28 weeks, and 34% (37/110) after 28 weeks. Almost half of the infants required over a day of mechanical invasive ventilation and prolonged parenteral nutrition, 46% (51/110) and 45% (49/110) respectively (**Table 1**).

*Table 1:* Description of the clinical population

| | n (%) | | | Mean (SD) |
|---|---|---|---|---|
| **Males/Females** | 59 (54) / 51 (46) | | | - |
| **Monochorionic twins** | 12 (11) | | | - |
| **Gestational age at birth: groups and mean (weeks)** | G1 (<26) 34 (31) | G2 (26-28) 39 (35) | G3 (>28) 37 (34) | 27.4 (2) |
| **SGA (<10th percentile)** | 23 (21) | | | - |
| **BPD** | 77 (70) | | | - |
| **Mechanical ventilation > 1 day** | 51 (46) | | | - |
| **NEC** | 15 (14) | | | - |
| **Parenteral nutrition ≥ 21 days** | 49 (45) | | | - |
| **Sepsis** | 68 (62) | | | - |
| **PMA at TEA-MRI (weeks)** | - | | | 41.1 (0.73) |

*Abbreviations: BPD: Bronchopulmonary dysplasia; MRI: magnetic resonance imaging; PMA: postmenstrual age; SGA: Small for gestational age; TEA: term equivalent age G1: <26w of gestational age (GA), G2: 26-28w GA, G3: 28-32w GA.*

The evaluation of preterm brains on TEA-MRI was first performed according to the Kidokoro scoring system. The intra-observer concordance coefficient Kappa for the whole brain score, based on repeated evaluations of 20 subjects, was 0.96 (95% CI:

0.90 – 1). Moderate to severe brain injuries were found in approximately 6% (7/110) of infants, while most were classified as normal (70%) (77/110) (**Table 2**). Anomalies were more frequent in grey matter and cerebellum than in white matter.

*Table 2:* *Distribution of cerebral injury severity according to Kidokoro scoring, hemorrhagic lesions distribution and DHESI frequency.*

| Kidokoro score: Severity | n (%) | | | |
|---|---|---|---|---|
| | **Normal** | **Mild** | **Moderate** | **Severe** |
| **Whole brain** | 77 (70) | 26 (23.6) | 4 (3.6) | 3 (2.7) |
| **Grey matter** | 71 (64.5) | 17 (15.5) | 18 (16.4) | 4 (3.6) |
| **White matter** | 86 (78.2) | 17 (15.5) | 3 (2.7) | 4 (3.6) |
| **Cerebellum** | 66 (60) | 21 (19.1) | 19 (17.3) | 4 (3.6) |
| **Hemorrhagic lesions*** | **SEH** | **IVH 2** | **IVH 3** | **IVH 4** |
| **39 (35.4)** | 14 (12.7) | 14 (12.7) | 6 (5.5) | 5 (4.5) |
| **DEHSI** | n (%) | | | |
| | 14 (13) | | | |

*\* According to Papile classification.*
*Abbreviations: IVH: Intraventricular hemorrhage (grades 2, 3, 4), SEH: Subependymal hemorrhage.*

Hemorrhagic lesions were identified in 35% of subjects. Note that significant hemorrhagic lesions were taken into account in the Kidokoro scoring either through the ventricular dilatation for grade III hemorrhages or as white matter signal anomaly in case of grade IV hemorrhagic lesions.

DEHSI (Diffuse excessive high signal intensity) was found in 13% of infants (**Table 2**). The quantification of brain compartment volumes revealed high variability across infants **(Figure 1b)**.

*A.*

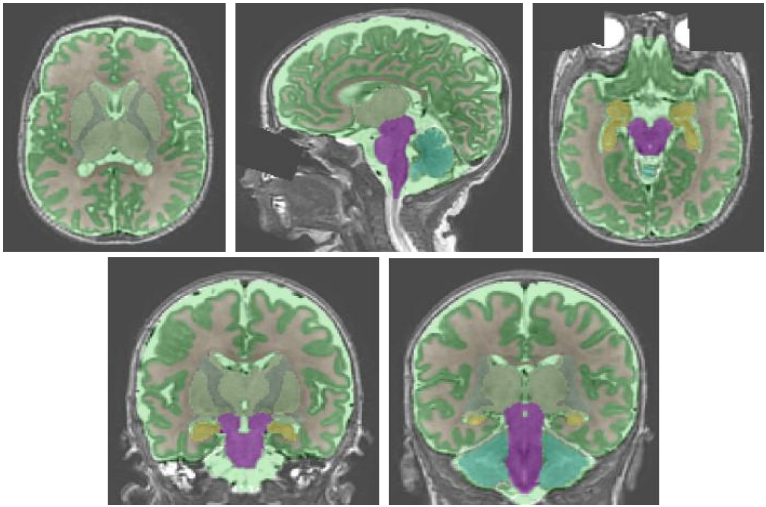

***Cortex*** – *Deep grey nuclei* – *subcortical white matter* – *central white matter* – *cerebellum* – ***brainstem*** – *hippocampus* – *cerebrospinal fluid*

*B.*

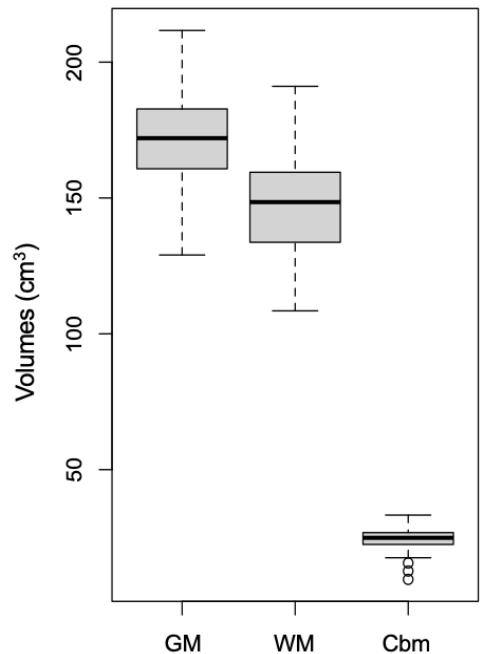

***Figure 1: A.*** *Segmentation of cerebral compartments according to a combination of iBEAT and DrawEM methods: cortex in emerald green, deep grey nuclei in sage green, subcortical white matter in white, central white matter (internal capsule) in grey, cerebellum in blue, brainstem in purple, hippocampus in yellow, cerebrospinal fluid in mint green.* ***B.*** *Box-Plot describing cerebral volumetry over the group of infants for the different brain compartments: grey matter GM (cortex and deep grey nuclei), white matter WM (subcortical and central) and cerebellum Cbm.*

**Clinical factors associated with injury severity according to Kidokoro scoring**

Multiple logistic regression analyses showed significant associations between clinical factors and degree of brain anomaly according to the Kidokoro scoring **(Figure 2)**.

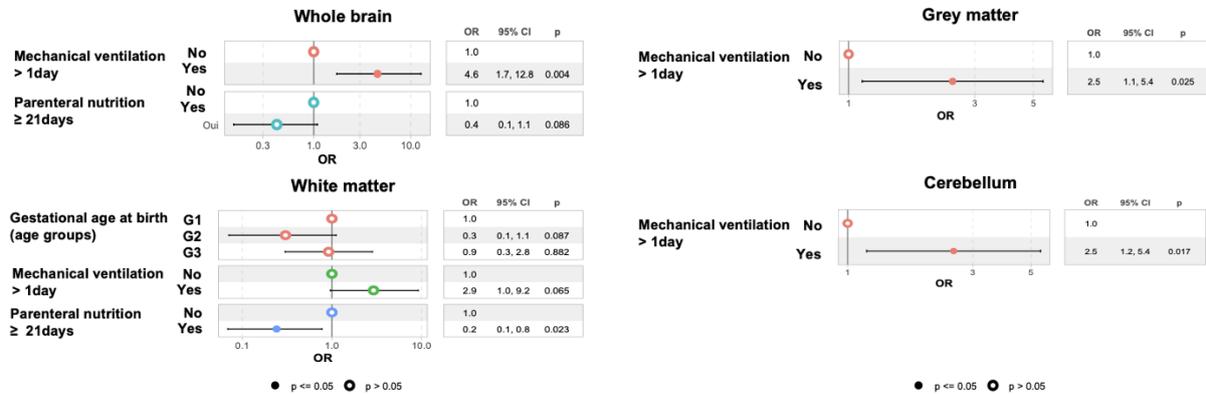

*Figure 2: **A.** Associations between clinical factors and injury severity according to Kidokoro score assessed with a logistic regression model (Odds-Ratio OR with their 95 % confidence interval CI) for the different brain compartments: **B.** Summary of the resulting significant associations and tendencies: \*\*\* p < 0.001 ; \*\* p < 0.01; \* p < 0.05; . p < 0.1*
*Abbreviations: GA: gestational age; G1: <26w GA, G2: 26-28w GA, G3: 28-32 GA; SGA: small for gestational age; BPD: bronchopulmonary dysplasia; NEC: necrotizing enterocolitis*

In the whole brain, we showed that a higher degree of Kidokoro injury severity was significantly associated with mechanical invasive ventilation (adjusted Odds Ratio aOR = 4.6, 95% Confidence Interval CI [1.7 – 12.8], p = 0.004), whereas we observed a trend for a lower degree of severity for infants who underwent prolonged parenteral nutrition (aOR = 0.4, 95% CI [0.1-1.1], p= 0.086).

Regarding the injury severity of grey matter, we found a similar association with mechanical invasive ventilation (aOR = 2.5, 95% CI [1.1 – 54], p = 0.025).

Regarding white matter, a similar trend was observed for mechanical invasive ventilation (aOR = 2.9, 95% CI [1.0 – 9.2], p = 0.065), whereas prolonged parenteral nutrition was identified as a significant protecting factor (aOR = 0.2, 95% CI [0.1 – 0.8], p = 0.023). The group with intermediate GA at birth (G2, birth between 26 and 28w GA) tended to show less injury severity (aOR = 0.3, 95% CI [0.1 – 1.1], p = 0.087).

In the cerebellum, we also found an increased injury severity in infants who had received mechanical invasive ventilation (aOR = 2.5, 95% CI [1.2 – 5.4], p = 0.017).

**Clinical factors associated with cerebral volumes**

Multiple linear regression was used to analyze cerebral volumes and their relationship with clinical factors, Kidokoro injury severity and PMA at TEA-MRI **(Figure 3).**

First, volumes of the whole brain, GM and cerebellum showed a significant increase with PMA at TEA-MRI, justifying the need for controlling analysis for this variable.

The global cerebral volume was significantly reduced when moderate or severe lesions were present according to Kidokoro scoring (p = 0.016 and 0.013 respectively). There was a trend for higher volumes in males (p = 0.05). Preterms born SGA (p < 0.001) or who had undergone mechanical invasive ventilation (p = 0.01) showed significantly decreased volumes.

Grey matter volume was significantly reduced in infants born as monochorionic twins (p = 0.028), SGA (p < 0.001), or having undergone mechanical invasive ventilation (p = 0.009). There was no association with brain injury severity according to Kidokoro scoring.

Concerning white matter volume, it was significantly reduced when moderate or severe lesions were present (p = 0.044 and 0.001 respectively), in infants born SGA (p < 0.001) or having undergone mechanical invasive ventilation (p = 0.002). Males had higher white matter volumes (p = 0.007).

Concerning cerebellar volume, it was significantly reduced in infants showing moderate and severe lesions (p = 0.003 and p < 0.001 respectively).

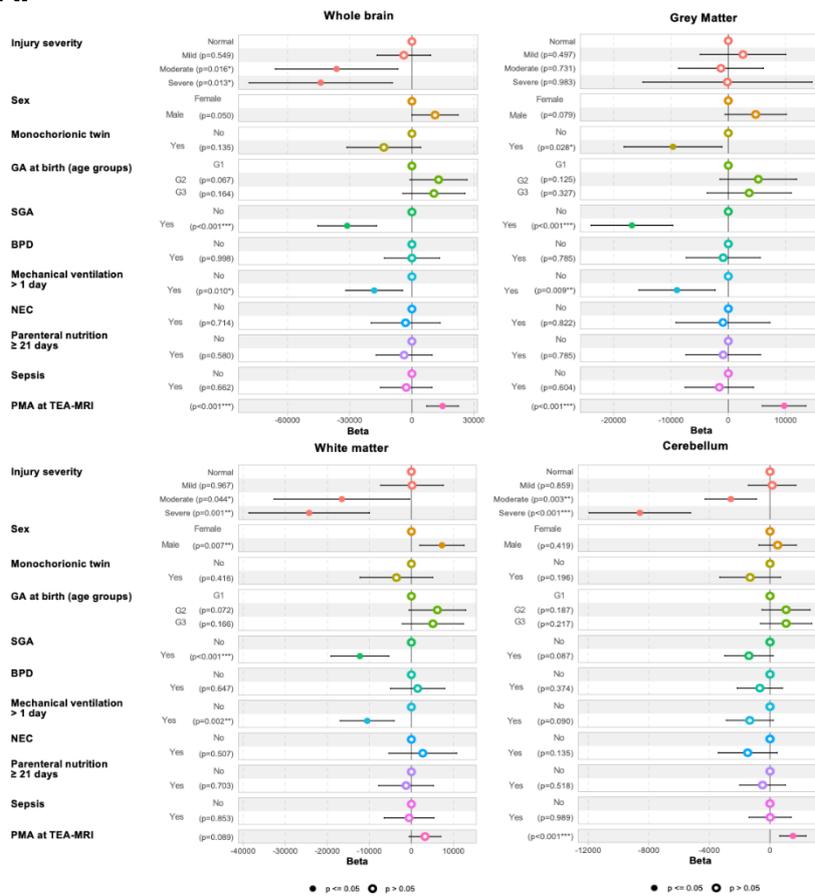

*Figure 3: **A.** Associations between cerebral volumetry, injury severity according to Kidokoro score, and clinical factors assessed according to a linear regression model for the different brain compartments: the whole brain, grey matter, white matter, cerebellum. **B.** Summary of the resulting significant associations and tendencies: \*\*\* p < 0.001; \*\* p < 0.01; \* p < 0.05; . p < 0.1*

*Abbreviations: GA: gestational age; G1: <26w GA, G2: 26-28w GA, G3: 28-32 GA; SGA: small for gestational age; BPD: bronchopulmonary dysplasia; NEC: necrotizing enterocolitis; PMA: postmenstrual age; MRI: magnetic resonance imaging; TEA: term equivalent age; GM: grey matter; WM: white matter*

## Discussion

In this study, we found mechanical invasive ventilation to be a strong risk factor for increased brain injury severity assessed with Kidokoro scoring in very and extreme preterm infants. Prolonged parenteral nutrition seemed to act as a protective factor on white matter abnormalities. Mechanical ventilation and being SGA at birth were also found to be strong risk factors for a reduction in cerebral volumes. There was a clear association between increased injury severity and reduced cerebral volumes.

A first striking aspect of this study was the distribution of brain injury severity we observed according to Kidokoro scoring compared to the one described in the original study by Kidokoro et al in 2013. They reported 35% of patients with moderate to severe injuries, whereas we only observed 6% [15]. Yet, this former study was set more than 10 years ago, and its population consisted of extremely premature newborns only. More recent studies revealed a distribution similar to ours, like the one from Abiramalatha et al. in 2022 with only 6% of patients having mild to severe injuries [27]. Still, this difference with the original study in injury severity and prevalence might explain a decreased significance of our results on relationships with clinical factors.

The association between mechanical ventilation and increased brain injury severity as well as reduced cerebral volumes has already been acknowledged in previous studies, such as the one from Brouwer et al. in 2017 [28]. This effect might be mediated through the cerebral inflammatory response to ventilation-induced oxidative stress [29], and / or the vasoconstrictive and ischemic effects of hypocarbia in the brain [30]. Our study further emphasizes the impact of mechanical ventilation on the early stages of cerebral development, but on the other hand we should keep in mind that this association might also reflect the dire clinical state of patients requiring such invasive ventilation.

Similarly, being SGA at birth was associated with reduced cerebral volumes (except for the cerebellum). Many infants born SGA suffered from vascular intra-uterine growth restriction (IUGR), which increases the risk of chronic ischemia in the brain and might therefore explain a volume reduction in those infants. Still, in previous research this reduction could be observed in all SGA preterms whether or not they had suffered from IUGR, implying other underlying mechanisms for this association [31] [32]. Interestingly, it was also shown that this cerebral volume reduction endures during infancy and adolescence, and is associated with the occurrence of neurodevelopmental disorders [33]. Nevertheless, since most infants born SGA do not recover a "normal" weight in the first post-natal weeks and stay small for PMA at term-equivalent age, one may wonder whether infants born SGA exhibited reduced cerebral volumes simply because of a lower weight at MRI, and not because of the reflection of cerebral lesions or global dysmaturation. However, we conducted a secondary analysis on 97 infants, controlling for the additional factor of being small for post-menstrual age at MRI, and confirmed the association between being born SGA and reduced cerebral volumes independently from small weight at MRI (see Supplementary Material).

Of note, male sex was associated with greater white matter volumes at TEA in our study. Still, this is most probably the consequence of sexual dimorphism, with boys exhibiting bigger cerebral volumes, the same way their birth weight and head circumference are higher than that of girls [34] [35] [36] [37]: this would explain why boys have bigger volumes despite female sex being a known protective factor in preterms.

Regarding the effect of GA at birth on brain injury severity, we observed a trend in the white matter, since the group with intermediate GA showed less often severe brain injury than the group with lowest GA. This was also observed in the original Kidokoro study [15] and by Brouwer et al. in 2017 [28]: the earlier the preterm is exposed to a possibly aggressive extra-uterine environment, the more the cerebral developmental course will be impacted. Still, this might not be the only explanation, and a higher degree of anomaly may exist from birth (or even before) in infants with lower gestational age: the only way to assess this hypothesis would be to collect early MRI data, before TEA, in those preterm babies. Yet the Kidokoro scoring system was not developed to assess such early MRI-scans. Of note, no difference was observed between the groups with lowest and highest GA, probably because this latter group was biased towards infants with brain abnormalities (not all infants born between 28 and 32 weeks GA were scanned, contrarily to infants born before 28 weeks GA).

Besides, no association was found between cerebral volumes, severity of anomalies and BPD, NEC, or sepsis, a surprising result considering that those factors are important mediators of inflammation that might lead to cerebral injury. BPD has been found to increase the risk of behavioral disorders and reduce IQ in preterms [38] [39]. Yet studies by Abiramalatha et al. [27] or Brouwer et al. [28] could not show an association with cerebral injury severity at TEA. This suggests that BPD impact might be mediated by anomalies that are not easily identified with MRI. Furthermore, although BPD and sepsis were identified as risk factors in previous studies [10], it is interesting to note that those pathologies were considered only in their most severe forms previously. Lee et al. did show reduced cerebral and white matter volumes in BPD infants, yet this association disappeared when only mild BPD were studied [40]. Balakrishan et al. [41] observed that sepsis increased injury severity, but they only considered culture-positive sepsis, while it has already been proven that culture-positive and culture-negative sepsis do not lead to similar neurodevelopmental outcomes. Finally, we considered NEC for all stages ≥ II, perforated and non-perforated, while it is known that patients with perforated NEC have higher odds of NDD than patients with non-perforated NEC [42]. Therefore, the lack of association in our study may be due to our choice in considering all kinds of BPD, both perforated and non-perforated NEC, and culture-negative sepsis in addition to culture-positive ones. Considering a larger cohort of infants would allow us to consider different levels for each factor, to be more precise and deal with such limitation.

A more surprising result was that prolonged parenteral nutrition seemed to act as a protective factor against cerebral injury, at least on white matter. Indeed, in previous studies a prolonged parenteral nutrition was associated with reduced biparietal perimeter in preterms and delayed neurodevelopment [43]. Yet those results are controversial, as an early parenteral nutrition was also found to be related to higher cranial perimeter in preterms [44]. We might suspect that the impact of parenteral nutrition varies depending on its composition or preparation technique: indeed, Costa et al. found an increase in cerebellar volumes when using a multicomponent lipid emulsion [45], and it was shown that the increased use of in-line filtration techniques for the preparation of parenteral nutrition could reduce their concentration in micro-particles and the associated inflammatory response [46]. The optimization of parenteral nutrition might be an interesting approach to limit cerebral injury in preterms and improve their neurodevelopmental prognosis.

In our research, we observed a higher number of risk factors and higher significance when cerebral volumes were analyzed rather than injury severity with the Kidokoro scoring. In the same way, the study by Nosaka et al. highlighted lower white matter volumes at TEA for preterms suffering from chorioamnionitis compared with controls but no difference in Kidokoro scores [11]. Analytic tools providing quantitative metrics such as cerebral volumes may be more sensitive for the detection of cerebral anomalies in preterms than qualitative ones like the Kidokoro scoring, and therefore would be more accurate to provide reproducible markers of prognosis. On the other hand, those quantitative analytic tools are not yet used in everyday practice, and qualitative ones are more accessible and still hold some prognostic value, as suggested by the association between decreased cerebral volumes and increased anomaly severity according to Kidokoro scoring.

Our study suffered from some limitations, like its retrospective nature, allowing only for restricted access to clinical data. Still, the infants being born in different hospitals, our population benefitted from a certain degree of diversity. Furthermore, having realized all MRI on the same scanner, we reduced the variability in our image quality. The main limitation in our study was actually the limited number of patients exhibiting moderate-to-severe cerebral injuries since this confers disproportionate weight to the clinical characteristics of these few subjects. Also, we could not find a proper index of clinical severity to use as a confounding factor in our multivariate analysis, therefore increasing the impact of factors such as mechanical ventilation. Finally, it should be noted that we did not consider in our analysis the use of certain medications frequently associated with mechanical ventilation, such as morphine or corticosteroids, known to decrease cerebral volumes and increase the risk of cerebral injury [47] [48]: this could act as a supplementary confounding factor. Indeed, we did not have access to the data regarding the cumulative dose of anesthetics received by each infant. And too few patients (2/110) had received strong doses of steroid to consider it as a factor in our statistic model.

Despite those limitations, the strength of our study is to offer both a qualitative and quantitative analysis of the preterm brain at TEA, relying on a standardized and optimized MRI protocol. The collected dataset on infants born very and extreme preterm will include neurodevelopment monitoring in the coming years, allowing us to further relate early neuroimaging results with outcome. This will also be the basis for future interventional studies, in which the risk and protective clinical factors that we identified could be used as confounding factors.

## Conclusion

This study provided an evaluation of the impact of clinical risk factors on early brain development in very and extreme preterms, emphasizing the negative impact of mechanical ventilation during the critical perinatal period, and demonstrating a protective role of parenteral nutrition. It further highlighted the benefits of using both a qualitative and quantitative analysis when exploring the preterm brain with TEA-MRI.

4.
[47]	M. M. Al-Mouqdad *et al.*, « Association between morphine exposure and impaired brain development on term-equivalent age brain magnetic resonance imaging in very preterm infants », *Sci Rep*, vol. 12, nº 1, p. 4498, mars 2022, doi: 10.1038/s41598-022-08677-0.
[48]	I. Robles, M. A. Eidsness, K. E. Travis, H. M. Feldman, et S. E. Dubner, « Effects of postnatal glucocorticoids on brain structure in preterm infants, a scoping review », *Neurosci Biobehav Rev*, vol. 145, p. 105034, févr. 2023, doi: 10.1016/j.neubiorev.2023.105034.

24